\documentclass[twocolumn,showpacs,preprintnumbers,amsmath,amssymb]{revtex4}

\usepackage{graphicx}
\usepackage{dcolumn}
\usepackage{bm}

\begin{document}

\preprint{APS/123-QED}

\title{Effect of charge state in nearby quantum dots on quantum Hall effect}%

\author{K. Takehana}
 \email{TAKEHANA.Kanji@nims.go.jp}
\author{T. Takamasu}
\author{G. Kido}%
\affiliation{%
Nanomaterials Laboratory, National Institute for Materials Science, 1-2-1 Sengen, Tsukuba, 305-0047 Japan
}%

\author{M. Henini}
\affiliation{
School of Physics and Astronomy, University of Nottingham, Nottingham, NG7 2RD, UK
}%

\date{\today}

\begin{abstract}
Magnetoresistance measurements have been performed on a gated two-dimensional electron system (2DES) separated by a thin barrier layer from a layer of InAs self-assembled quantum dots (QDs). 
Clear features of the quantum Hall effect were observed despite the proximity of the QDs layer to the 2DES. 
However, the magnetoresistance ($\rho_{xx}$) and Hall resistance ($\rho_{xy}$) are suppressed significantly in the magnetic field range of filling factor
 $\nu < 1$ when a positive voltage is applied to the front gate. 
The influence of the charge state in QDs was observed on the transport properties of the nearby 2DES only in the field range of $\nu < 1$. 
It is proposed that the anomalous suppression of $\rho_{xx}$ and $\rho_{xy}$ is related to spin excitation, which is induced by spin-flip processes involving electrons in the QDs and the 2DES.
\end{abstract}

\maketitle

\section{\label{sec_intro}Introduction}
Semiconductor quantum dots (QDs), especially those formed by the Stranski-Krastanow growth mode, have attracted great deal of attention, due to their promising potential for applications in electronic memories, optoelectronic devices, etc. 
Many studies have focused on applications of QDs floating gates, that is, to manipulate electrons in the QDs by using light or electrostatic gating and to detect the electron state in QDs by measuring the transport properties of the nearby two-dimensional conducting channel \cite{Sakaki01,Q_Wang,TH_Wang,Schliemann,Koike01}. 
Sakaki {\it et al.} found that InAs QDs strongly affected the transport properties of a nearby two-dimensional electron system (2DES) \cite{Sakaki01}. 
Well-defined large hysteresis and memory effects were observed in the gate voltage dependence of the conductance of a narrow channel field-effect transistor (FET) \cite{Schliemann,Koike01}. 
These properties were assigned to the effect of charging and discharging of QDs on the electron scattering processes in the 2DES. Furthermore, single photon detection due to discharging of individual QDs has been demonstrated on a similarly designed FET device \cite{Shields}.  

It is also well-known that the quantum Hall effect (QHE) appears in a high quality 2DES at sufficient low temperature in the presence of a strong magnetic field and several investigations have been performed in order to study the influence of a layer of QDs on the transport properties of a nearby 2DES in the quantum Hall regime \cite{Ribeiro,Q_Wang,Ribeiro01,GH_Kim}. 
Clear features of Shubnikov-de Haas (S-dH) oscillations and the integer QHE were observed in samples with high mobility, in spite of the presence of a nearby QD layer \cite{Ribeiro,Q_Wang}. 
Coulomb scattering by electrons in charged QDs was found to play an important role in the edge state transport \cite{Q_Wang}. 
A QHE-insulator transition was found when a high magnetic field was applied on a 2DES incorporating QDs \cite{Ribeiro01,GH_Kim}. 
It was argued that QDs cause the strong scattering of electrons in the 2DES, which induces the QHE-insulator transition. 
Wang {\it et al.} investigated the influence of QDs on the transport properties of a nearby 2DES in an (InGa)As/InP quantum well (QW) in the presence of high magnetic fields, and found the $``$overshoot$"$ effect in the quantum Hall regime when the QDs are charged with electrons \cite{Q_Wang}. 
The $``$overshoot$"$ effect was interpreted as results in the coupling between spin-split edge states, which indicates the presence of spin-flip processes \cite{Komiyama,Richter}.
However, no experimental evidence has been obtained concerning with the correlation between the electron states in the QDs and those of 2DES. 
We expect that additional scatterings and/or other effects will appear in the 2DES in the quantum Hall regime, due to the correlation with the electron charging and/or spin states localized in nearby QDs.

In this study, we have measured the magnetoresistance and Hall resistance of a 2DES separated by a thin barrier from a layer of InAs QDs in high magnetic fields and at low temperature ($T = 0.5$ K), in order to investigate the influence of the electron charge state of the QDs on the transport properties of the nearby 2DES.

\section{Experimental Detailes}

\begin{figure}
\includegraphics*[width=7cm]{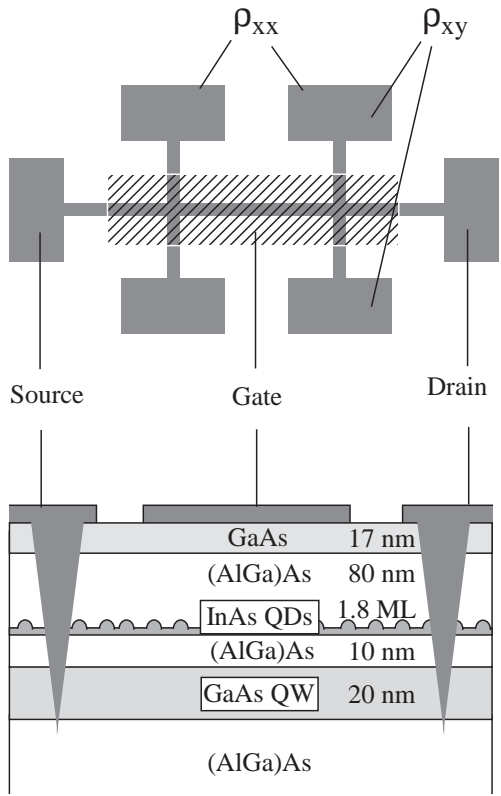}%
\caption{\label{fig1} Structure of the devices used in this study.}
\end{figure}

Figure~\ref{fig1} shows schematically the structure of the devices used in this study. 
Samples were prepared from a layer grown by molecular beam epitaxy (MBE) on a semi-insulating GaAs substrate. 
The 2DES channel is formed as a 20 nm wide well in a modulation-doped (AlGa)As/GaAs/(AlGa)As heterostructure. 
The InAs QDs layer is embedded in an (AlGa)As barrier and is separated from the 2DES channel by 10 nm (AlGa)As layer. 
From atomic force microscope (AFM) measurements on a reference sample, the diameter, height and density of the QDs are estimated to be 30 nm, 3 nm and 1 $\times$ 10$^{11}$ cm$^{-2}$, respectively. 
There are two Si-doped AlGaAs layers to provide modulation doping of the well; one is located between the substrate and 2DES, the other is between QDs and surface. 
Samples were fabricated in the form of standard Hall bars of 10 $\mu$m width. 
A semi-transparent front gate was deposited on the Hall bar, in order to modulate the carrier density. The gate voltage also acts to vary the energy levels of the QDs with respect to the Fermi energy of the 2DES. 
Transport and capacitance measurements were performed in a 15 T superconducting magnet with $^3$He refrigerator, in which the sample was cooled down to $T$ = 0.5 K.

\section{\label{sec_data}Experimental Results}

\begin{figure}
\includegraphics*[width=8cm]{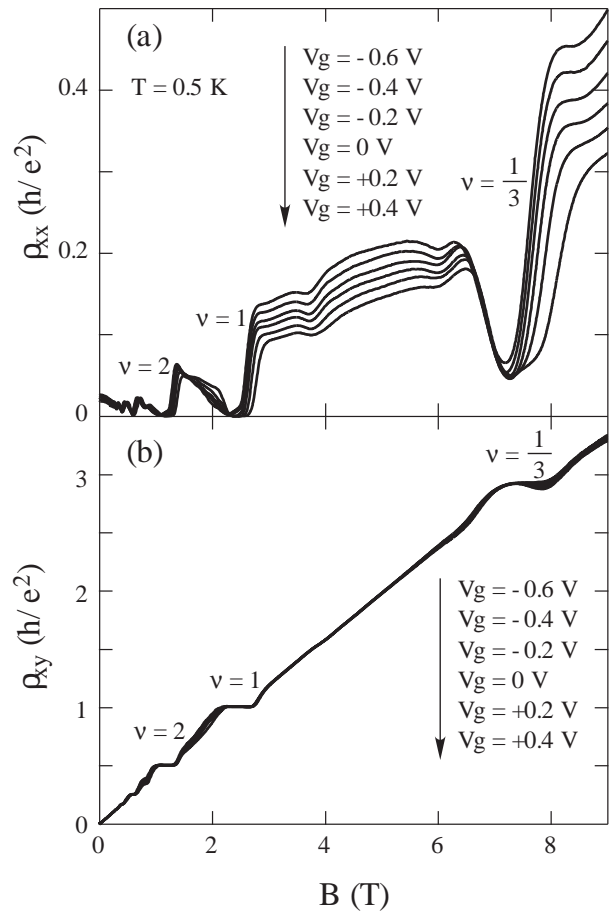}
\caption{\label{fig2} Magnetic field dependence of the (a) longitudinal ($\rho_{xx}$) and (b) Hall ($\rho_{xy}$) magneto-resistivity of the device at 0.5 K, respectively, with various applied gate voltage, before illumination of light.}
\end{figure}

The traces in Fig.~\ref{fig2}(a) and (b) show the longitudinal ($\rho_{xx}$) and Hall ($\rho_{xy}$) resistivity of the device measured at $T = 0.5$ K as a function of magnetic field and at various applied gate voltages $V_g$. 
The $\rho_{xx}$ and $\rho_{xy}$ show clear features of both integer and fractional QHE, indicating a relatively high quality 2DES, in spite of proximity of the QDs. 
The S-dH oscillations can be seen in $\rho_{xx}$ up to $\nu = 12$ in the low magnetic field range.
The carrier density ($n$) and mobility ($\mu$) of the device are found to be the $n$ = 5 $\times$ 10$^{10}$ cm$^{-2}$ and $\mu = 2 \times 10^5$ cm$^2$/Vs, respectively, at $V_g$ = 0 V. 
When the gate voltage is changed, the transport properties are affected mainly by the change of the carrier density in 2DES. 
The mobility also increases with the carrier density. 
The gate voltage dependence of the transport properties shown in Fig.~\ref{fig2} can be understood within the standard QHE model of the 2DES whose carrier density is modulated by the applied gate voltage.

The gate voltage dependence of the carrier density is, however, much smaller than the expected value deduced from the geometric capacitance by neglecting presence of QDs and Si-doping layer. 
There are two possible origins for such a small dependence; one is the presence of QDs, and the other is Si-doping layer. 
As for the former possibility, an additional capacitance, the so-called $``$quantum capacitance$"$, should be considered in addition to the geometric capacitance between the QDs and 2DES, because the QDs are located close to the 2DES \cite{Ando,Luryi,Macucci}.  
In this case, however, the $``$quantum capacitance$"$ of QDs is comparable with the geometric one, and is not large enough to explain experimental results. 
As for the later possibility, the density of Si donors in the modulation-doped layer between the 2DEG and surface is $3 \times 10^{12}$ cm$^{-2}$ in this sample. 
It is relatively high, and pinning of the Fermi energy can occur at the Si donor level. 
When the Fermi energy is pinned at the donor level, the gate voltage does not affect the carrier density in 2DES as much as estimated geometrically, which can explain the experimental results. 
Therefore, the small gate voltage dependence of the carrier density is attributed to pinning of the Fermi energy at the Si doping level.

\begin{figure}
\includegraphics*[width=8cm]{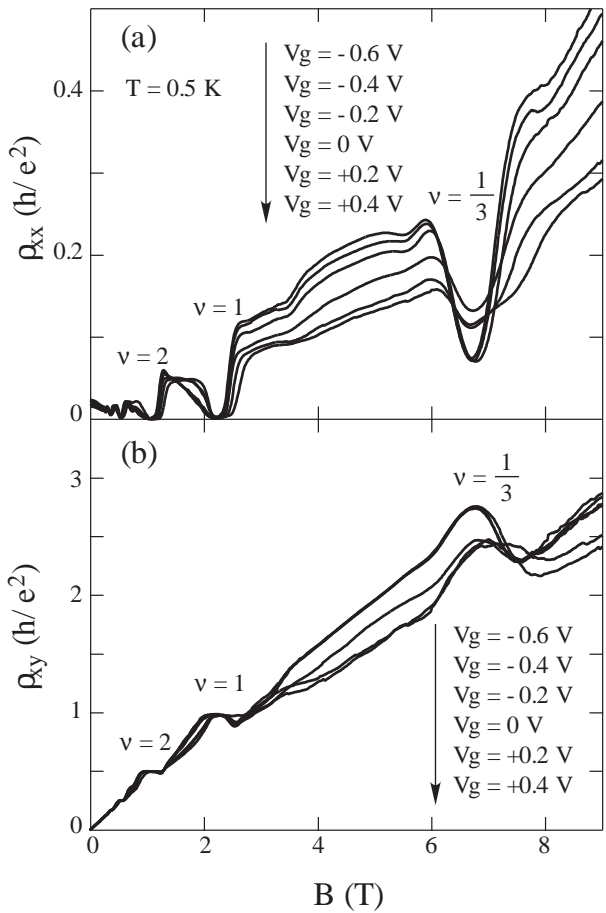}
\caption{\label{fig3} Magnetic field dependence of the (a) $\rho_{xx}$ and (b) $\rho_{xx}$ of the device at 0.5 K, respectively, with various applied gate voltage, after illumination of weak light.}
\end{figure}

After illumination with a low level of above-bandgap light, the gate voltage dependence of the transport property is changed drastically. 
Figures~\ref{fig3}(a) and (b) show the gate voltage dependence of $\rho_{xx}$ and $\rho_{xy}$, respectively. 
Both magnetic and gate voltage dependence of $\rho_{xx}$ and $\rho_{xy}$ for $V_g < -0.2$ V are quite similar to those before illumination, 
while the plateau at $\nu = \frac{1}{3}$ in $\rho_{xy}$ changes to a maximum, and a minimum appears on the higher field side. 
A small minimum of $\rho_{xy}$ was also observed on the higher field side of the plateau of $\nu = 1$ and $\nu = 2$.
The gate voltage dependence of the carrier density is also as small as before illumination, and does not show a significant anomaly around $V_g = 0$ V. 

For $V_g > 0$ V, the traces of $\rho_{xx}$ show S-dH oscillations in the lower field range, and the plateaus of $\rho_{xy}$ for the integer QHE are in good agreements with $\frac{h}{\nu e^2}$, as well as in the case of before illumination. 
Parallel conduction effect are probably small, because $\rho_{xy}$ is consistent with the theoretical value for integer QHE and $\rho_{xx}$ is almost zero at $\nu = 1$ and 2. 
In the field range of $\nu < 1$, however, the transport properties, especially the behavior of $\rho_{xy}$, is significantly different from that before illumination. 
The value of $\rho_{xy}$ is suppressed considerably from the theoretical value of $\frac{B}{ne}$ in the field range of $\nu < 1$. 
The deviation increases with applied magnetic field, while it becomes smaller in the vicinity of $\nu = \frac{1}{3}$. 
The value of $\rho_{xx}$ is also suppressed compared with the values for $V_g < -0.2$ V. 
The traces of both $\rho_{xx}$ and $\rho_{xy}$ become noisy in the restricted field range of $\nu < 1$ and for $V_g > 0$ V.

\begin{figure}
\includegraphics*[width=8cm]{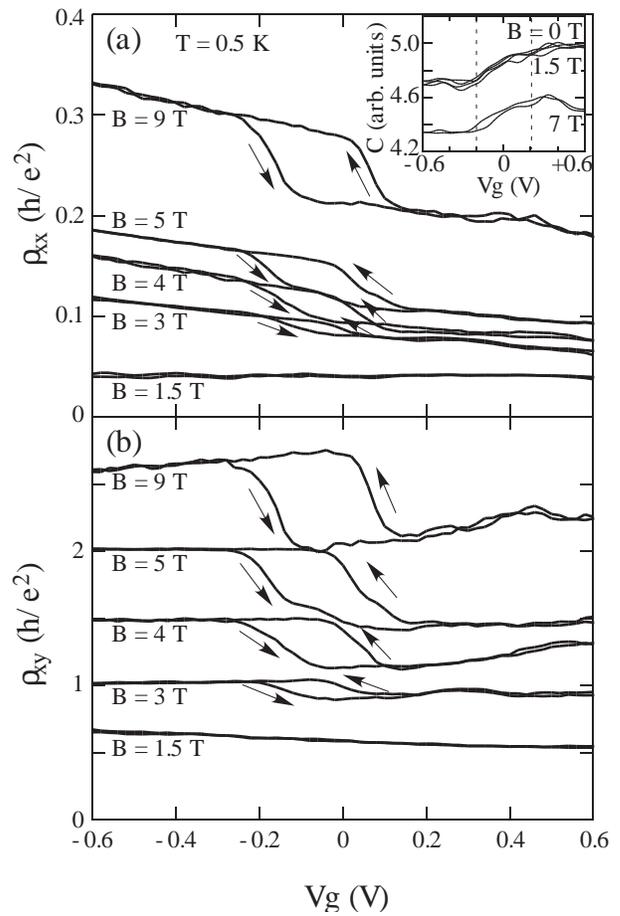}
\caption{\label{fig4} Gate voltage dependence of (a) $\rho_{xy}$ and (b) $\rho_{xy}$, respectively, with fixed magnetic field , after illumination of weak light. 
Traces of the inset show the gate voltage dependence of the capacitance between the front gate and 2DES at $T = 0.5$ K.}
\end{figure}

In order to make this anomalous behavior clear, the gate voltage dependence of $\rho_{xx}$ and $\rho_{xy}$ was measured with fixed magnetic field. 
The traces of $\rho_{xx}$ and $\rho_{xy}$ versus $V_g$ are shown in Fig.~\ref{fig4}(a) and (b), respectively. The gate voltage was swept from -0.6 to +0.6 V and back to -0.6 V. 
At magnetic field below $\nu = 1$, $\rho_{xx}$ and $\rho_{xy}$ change with $V_g$ smoothly and show no anomaly around $V_g = 0$ V. 
In the field range above 3 T, on the other hand, both $\rho_{xx}$ and $\rho_{xy}$ steeply decrease around $V_g = -0.2$ V on the $V_g$-upsweep, and steeply increase around $V_g = +0.2$ V on the $V_g$-downsweep. 
As a result, a well-defined hysteresis loop was observed at the narrow gate voltage range between -0.2 and +0.2 V on the traces of $\rho_{xx}$ and $\rho_{xy}$. 
Both thresholds of the hysteresis loop are almost independent of the applied magnetic field. 
The magnetic fields where the hysteresis loop was observed correspond to $\nu < 1$. 
Both $\rho_{xx}$ and $\rho_{xy}$ exhibit higher resistance on the lower gate voltage side of the hysteresis loop, and a lower resistance on the higher gate voltage side. 
The transport property seems to switch between higher and lower resistance mode by sweeping the gate voltage within narrow voltage range of $\sim$0.2 V. 
The transition between the higher and the lower resistance has a time constant $>$ 60 s around the thresholds. 
The value of $\rho_{xy}$ is suppressed from the theoretical value of $\frac{B}{ne}$ in the low resistance mode, while it is comparable with that value in the high resistance mode. 
The difference of both $\rho_{xx}$ and $\rho_{xy}$ between the high and low resistance modes, $\Delta\rho_{xx}$ and $\Delta\rho_{xy}$, increases with the applied magnetic field at $\nu <1$, 
suggesting that the anomalous suppression of $\rho_{xx}$ and $\rho_{xy}$ have magnetic origin. 
Note that the suppression ratios, $\frac{\Delta\rho_{xx}}{\rho_{xx}}$ and $\frac{\Delta\rho_{xy}}{\rho_{xy}}$, are similar to each other. 
For instance, $\frac{\Delta\rho_{xx}}{\rho_{xx}} \sim \frac{\Delta\rho_{xy}}{\rho_{xy}} \sim 0.2$ at $B = 6$ T. 
The $\Delta\rho_{xx}$ and $\Delta\rho_{xy}$ vanished at around $\nu \sim 1$. 

The capacitance, $C$, between 2DES and the front gate was also measured as a function of the gate voltage with fixed magnetic field, as shown in the inset of Fig.~\ref{fig4}. 
The experimental results of capacitance measurements exhibit a gate voltage dependence in the same range of -0.2 V $< V_g < +0.2$ V as the hysteresis loop and of $\rho_{xx}$ and $\rho_{xy}$ observed, though no hysteresis observed. 
It is notable that the capacitance change versus the gate voltage was observed not only in the field range of $\nu < 1$, but also in the field range of $\nu > 1$.

\section{Discussion}

\begin{figure}
\includegraphics*[width=8cm]{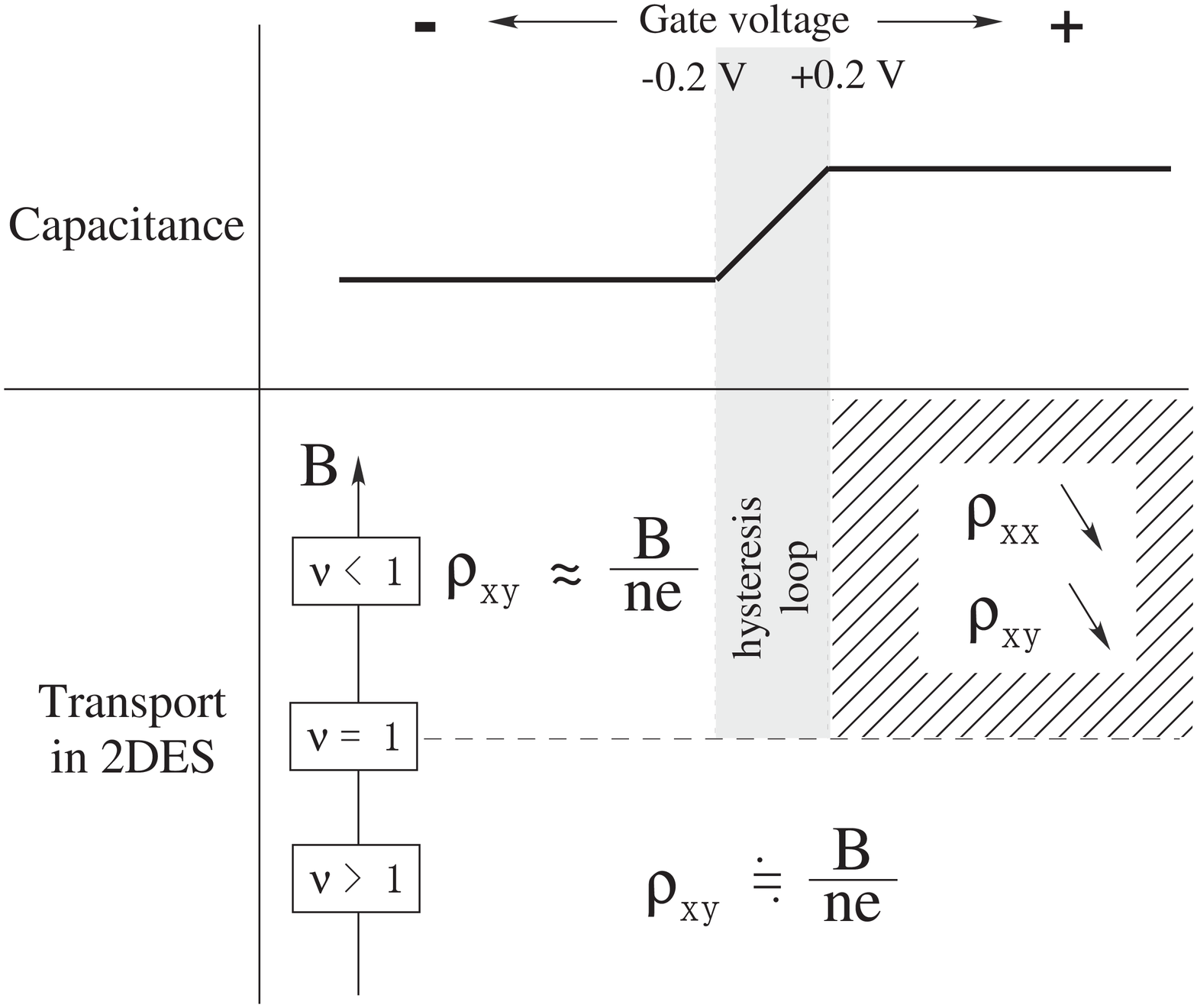}
\caption{\label{fig5} Summary of the experimental results of the capacitance and transport measurements.}
\end{figure}

In Fig.~\ref{fig5}, we summarize the experimental results of the capacitance and transport measurements after illumination of light. 
A significant change was observed in the gate voltage range between $V_g = -0.2$ V and +0.2 V in both measurements, while the anomalous hysteresis loop was observed only in the field range of $\nu < 1$ in the transport measurements. 
Recent measurements on the narrow channel devices of the similar structures revealed a hysteresis loop on the gate voltage dependence of the conductance, attributed to charging and discharging of QDs \cite{Schliemann,Koike01}. 
This fact assures that the applied gate voltage changes the number of electrons in QDs, which, in turn, affects the transport property of nearby 2DES. 
A clear feature has also been reported in the capacitance measurements on a similar sample, when the number of electrons in QDs is changed \cite{Miller,Medeiros}. 
In our study, the significant structure was found in the same gate voltage region, which is accompanied by the hysteresis loop on the transport measurements. 
From these facts, we assigned that charging and discharging occur in the QD in the gate voltage range between $V_g = -0.2$ V and +0.2 V. 
Considering the geometric capacitance of the device, the charge induced by sweeping the gate voltage over $\Delta V_g \sim 0.4$ V, which is the width of the hysteresis loop, is estimated to be 10$^{11}$ cm$^{-2}$. 
This value coincides with the density of QDs, indicating one-electron charging or discharging in this transition region.

Next, we have to discuss about how the electron state of the QDs affects the transport properties of the 2DES. 
The presence of the QDs can affect the 2DES through two types of scattering; one is short range scattering induced by fluctuation of alloy composition or by strain, and the other is Coulomb scattering arising from any change in the QDs. 
The former must be almost independent of the applied magnetic field. 
In the case of Coulomb scattering, only the exchange term should be strongly affected by the application of magnetic field. 
Thus, the spin-related interaction should play an important role in the results of the present study. 
In recent study, the influence of the electron charge state in QDs was investigated on the transport properties of a nearby 2DES in an (InGa)As/InP QW in the quantum Hall regime \cite{Q_Wang}. 
The $``$overshoot$"$ effect was observed, owing to charging effect of the QDs at odd filling factor. 
It indicates the coupling between spin-split edge states, which indicates the presence of spin-flip processes \cite{Komiyama,Richter}.
In the present case, the significant suppression of $\rho_{xx}$ and $\rho_{xy}$ was found when the QDs are charged with electrons only at $\nu < 1$ where the electron spin is fully polarized. 
The total spin state of artificial atom of the QDs is changed by charging and discharging, which means that the spin state of QDs at $V_g < 0$ V is different from that at $V_g > 0$ V. 
We infer that the reduction of $\rho_{xx}$ and $\rho_{xy}$ originates in the spin-flip processes involving the spin of the electrons in the QDs and the free carrier spins of the 2DES. 
Such spin-flip processes can occur, because Zeeman splitting of the QDs states is comparable to that of the carriers in the 2DES of the GaAs QW. 
The effective $g$-value, $g*$, of the InAs QDs has been reported to be $|g*| = 0.7 \sim 1.6$ \cite{Itskevich,JM_Meyer,Medeiros02}. 
The $``$overshoot$"$ effect was not observed, because the integer QHE is not clearly observed at odd filling factor in our study.

\begin{figure}
\includegraphics*[width=8cm]{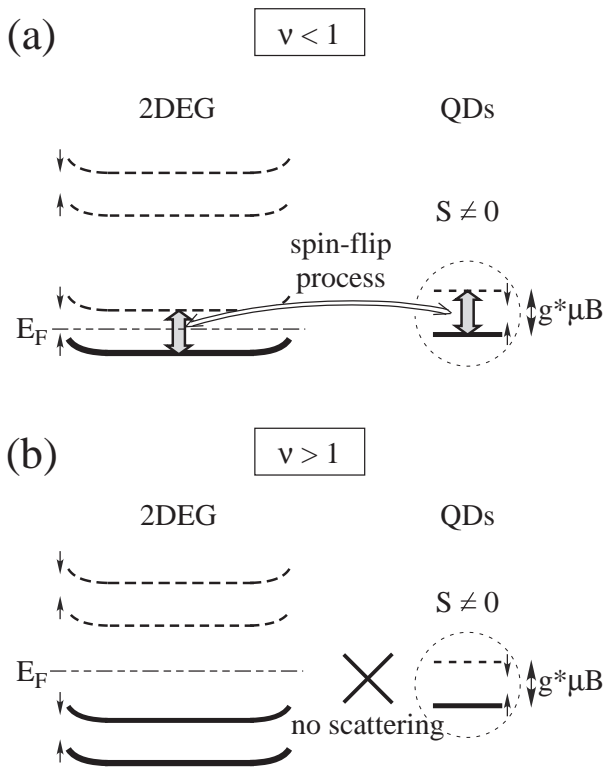}
\caption{\label{fig6} Schematic explanation for spin excitation induced by spin-flip process. 
(a) Spin excitation in 2DES across the Zeeman splitting can be induced by spin-flip process between 2DEG and QDs at $\nu < 1$, because Zeeman splitting in QDs is comparable with that of carrier of 2DES in GaAs QW. 
(b) Excitation between Landau levels does not occur $\nu >$ 1.
}
\end{figure}

We now consider the mechanism of reduction of $\rho_{xx}$ and $\rho_{xy}$ when spin-flip processes occur involving the spin of the QDs and the carrier spins of the 2DES. 
It is notable that both $\rho_{xx}$ and $\rho_{xy}$ are reduced by a similar degree on the higher gate voltage side in the field range of $\nu < 1$, see Sec.~\ref{sec_data}. 
In the classical approximation, $\rho_{xx}$ and $\rho_{xy}$ are given by $\frac{m*}{ne^2\tau}$ and $\frac{B}{ne}$, respectively. 
This suggests that the reduction of $\rho_{xx}$ and $\rho_{xy}$ by the same degree in the hysteresis step around $V_g = 0$ V arises from an increase in the carrier density. 
However, no anomalous change was observed concerning with the carrier density in 2DES on the gate voltage region where $\rho_{xx}$ and $\rho_{xy}$ are reduced. 
To explain this apparent increase in the carrier density, we propose that the reduction of $\rho_{xx}$ and $\rho_{xy}$ is induced by spin excitation due to spin-flip processes between the 2DES and the QDs, as indicated in Fig.~\ref{fig6}(a). 
When spin-flip processes occur, electrons will be excited to the down-spin Landau level and behave as additional electrons. 
Because remained quasi-holes in the up-spin Landau level immediately rearrange and recombination of the generated pair need long relaxation time \cite{Komiyama02}, this $``$hot-electron$"$ state continues long time. 
The coupling between edge states of up-spin and down-spin Landau levels might occur due to the spin-flip processes. 
If the excited electrons contribute to the transport in 2DEG, it induce the reduction of $\rho_{xx}$ and $\rho_{xy}$ by same degree, while the $``$real$"$ carrier density does not change. 
The electrical noise accompanying the anomalous reduction of $\rho_{xx}$ and $\rho_{xy}$ indicates the presence of the non-equilibrium state in 2DEG. 
The increase of the spin-flip process acts as increase of the $``$effective$"$ carrier density, and the reduction of $\rho_{xx}$ and $\rho_{xy}$. 
This explanation is supported by the fact that $\Delta\rho_{xx}$ and $\Delta\rho_{xy}$ increase with the magnetic field, 
because Zeeman splitting is proportional to the magnetic field. 
On the other hand, since the energy gap between Landau levels in 2DES is much larger than the Zeeman splitting, no spin excitation will occur when the filling factor $\nu$ is even (see Fig.~\ref{fig6}(b)). 
The spin excitation can also occur when filling factor $\nu$ is odd, but it is not significant in our data, because the integer QHE for odd filling factors is not clearly observed, due to the insufficient splitting of spin gap at $\nu > 3$. 
Therefore, the spin-flip process modulates the spin and electron ground state of the QHE only in the field range of $\nu < 1$, while it does not modulate the carrier density in 2DES.

Finally, we note the effect of light illumination. 
Prior to illumination, no anomaly is observed in the gate voltage dependence. 
Following a low level of illumination, however, an anomalous reduction of $\rho_{xx}$ and $\rho_{xy}$ occurs on the higher gate voltage side in the field range of $\nu < 1$. 
In this device, pinning of the Fermi energy is likely to occur at the Si-donor level located between the gate and the 2DEG, because the gate voltage dependence of the carrier density is much smaller than that estimated by the geometric capacitance, as discussed in Sec.~\ref{sec_data}. 
Illumination by a low level of light could excite electrons in Si-donor level and ionize the Si donor states. 
This would tend to reduce the pinning of the Fermi energy and modulate the local band bending \cite{G_Li}. 
We infer that charging and discharging of QDs can occur by applying gate voltage due to the effect of illumination, which affects the transport properties of 2DES as discussed above. 
The DX centers and/or other Si localized states might be concerned with this phenomenon, while the detailed mechanism is not clear at present. 

In summary, we have measured the transport and properties on 2DES incorporating a layer of InAs self-assembled QDs separated by a thin barrier layer with applied gate voltage in the presence of high magnetic fields, in order to investigate the influence of the electron state in QDs on the transport properties of the nearby 2DES. 
Clear features of integer and fractional QHE were observed despite the presence of a QDs layer close to the 2DES. 
Significant suppression of $\rho_{xx}$ and $\rho_{xy}$, however, was observed only in the field range of $\nu < 1$ and $V_g > 0$ V. 
We propose that charging and discharging of QDs occurs due to the applied gate voltage around $V_g \approx 0$ V. 
The electron charge states of the QDs affect the transport properties of the nearby 2DES only in the field range of $\nu < 1$. 
The anomalous suppression of $\rho_{xx}$ and $\rho_{xy}$ can be induced by spin excitation, owing to spin-flip process between QDs and carriers in the 2DES.

\begin{acknowledgments}
Dr.~M. Henini acknowledges the support from the Engineering and Physical Sciences Research Council (UK) and the SANDiE Network of Excellence of the European Commission, contract number NMP4-CT-2004-500101.
We are grateful to Professor~L. Eaves for useful discussion. 
\end{acknowledgments}

\end{document}